\documentclass[12pt,preprint]{aastex}

\shorttitle{NLS1s on the $M_{\rm BH}-\sigma$ relation}
\shortauthors{S. Komossa \& D. Xu}

\begin{document}

\title{Narrow-line Seyfert 1 Galaxies and the $M_{\rm BH} - \sigma$ Relation}

\author{S. Komossa}
\affil{Max-Planck-Institut f\"ur extraterrestrische Physik,
Postfach 1312, 85741 Garching, Germany; skomossa@mpe.mpg.de}

\and

\author{D. Xu}
\affil{National Astronomical Observatories, Chinese Academy of Science,
A20 Datun Road, Chaoyang District,
Beijing 100012, China; 
dwxu@bao.ac.cn}

\begin{abstract}
We have studied the location of narrow-line Seyfert 1 (NLS1) galaxies
and broad-line Seyfert 1 (BLS1) galaxies on the 
$M_{\rm BH} - \sigma$ relation of non-active galaxies.
We find that NLS1 galaxies as a class
 -- as well as the BLS1 galaxies
of our comparison sample -- {\em do follow} the
$M_{\rm BH} - \sigma$ relation of non-active galaxies 
if we use the width of the [SII]6716,6731 emission lines
as surrogate for stellar velocity dispersion, $\sigma_*$.
We also find that the width of [OIII]5007 is a good surrogate
for $\sigma_*$, but only after (a) removal of asymmetric blue wings,
{\em and}, more important, after (b) excluding core [OIII] lines 
with strong blueshifts
(i.e., excluding galaxies which have their
[OIII] velocity fields dominated by radial motions, presumably outflows).
The same galaxies which are extreme outliers in [OIII] 
still follow the $M_{\rm BH} - \sigma$ relation in [SII].  
We confirm previous findings 
that NLS1 galaxies are systematically off-set from the 
$M_{\rm BH} - \sigma$ relation if the full [OIII] profile
is used to measure $\sigma$. 
We systematically investigate
the influence of several parameters
on the NSL1 galaxies' location on the $M_{\rm BH} - \sigma$ plane:
[OIII]$_{\rm core}$ blueshift, $L/L_{\rm Edd}$,
intensity ratio FeII/H$\beta$, NLR density, and absolute magnitude.
Implications for NLS1 models
and for their evolution along the $M_{\rm BH} - \sigma$ relation
are discussed.
\end{abstract}


\section{Introduction}

Correlations between the properties of 
galaxies and the masses of their central black holes
are of fundamental importance for our understanding of 
galaxy formation and evolution. 
The relations
between black hole mass and bulge stellar velocity
dispersion, $M_{\rm BH} - \sigma_*$,
(Ferrarese \& Merritt 2000,
Gebhardt et al. 2000, Merritt \& Ferrarese 2001)
and between black hole mass and bulge mass (Marconi \& Hunt 2003, H{\"a}ring \& Rix 2004)
imply a close link between black hole and bulge formation and growth
(e.g., Haehnelt \& Kauffmann 2000). 

Mathur et al. (2001) emphasized the importance 
of studying the locus of NLS1 galaxies on
the $M_{\rm BH} - \sigma$ plane. 
NLS1s are a subclass of Active Galactic
Nuclei (AGN) characterized
by small widths of their broad Balmer lines
(e.g., Veron-Cetty et al. 2001). They are believed 
to have low-mass black holes (BHs) accreting close to the Eddington
limit (e.g., Boroson 2002).
As such, they are an 'extreme' class of AGN in
the local universe, potentially
holding important clues as to the origin of the
$M_{\rm BH} - \sigma$ relation
(see Sect. 1 of Li et al. 2006
for a review on models), and whether it holds
at all times and for all types of galaxies,
and at low BH masses.

However, previous studies of NLS1 galaxies, employing different
samples and methods, led to
partially conflicting results regarding the galaxies' location
on the $M_{\rm BH} - \sigma$ relation.
When using the width of the narrow [OIII]5007
emission line, FWHM$_{\rm [OIII]}$, as replacement
for stellar velocity dispersion $\sigma_*$,
most authors found that, on average,
NLS1 galaxies are {\em off} the $M_{\rm BH} - \sigma_{\rm[OIII]}$
relation (Mathur et al.
2001, Grupe \& Mathur 2004, Mathur \& Grupe
2005a,b, Bian et al. 2006, Watson et al. 2007), while in the study of
Wang \& Lu (2001) most NLS1s are, on average, 
{\em on} the relation
(with a significant scatter in all cases).
Furthermore, NLS1 galaxies were found to be off (Wandel 2002, Ryan et al. 2007)
or on (Botte et al. 2004) the
black-hole-mass -- bulge-luminosity, $M_{\rm BH} - L_{\rm bulge}$,
relation; and on (Botte et al. 2005, 
Barth et al. 2005)
or off (Zhou et al. 2006) the
$M_{\rm BH} - \sigma_{*}$ relation.
Most larger samples 
are actually based on 
FWHM$_{\rm [OIII]}$/2.35 as surrogate for $\sigma_*$ (Nelson 2000),
because the bulge velocity
dispersion $\sigma_*$ is very difficult to measure
in NLS1 galaxies due to superposed emission-line complexes.

Given the potentially far-reaching consequences
of a deviation of low BH mass, highly accreting
AGN from the $M_{\rm BH} - \sigma$ relation, 
independent studies of large(r) samples of
NLS1 galaxies are important.
No matter whether they turn out
to be on or off the $M_{\rm BH} - \sigma$ relation
of non-active galaxies and BLS1 galaxies, 
any result has important consequences: 
if NLS1s lie off the relation, 
for the formation of the $M_{\rm BH} - \sigma$ relation
(how do objects
move onto the relation ?)
and/or the gaseous kinematics
of their NLRs. If they lie on the relation the question
is raised how to preserve $M_{\rm BH} - \sigma$
{\em along} the relation  in objects which
are rapidly growing their BHs but are typically in non-merging
galaxies (Krongold et al. 2001).
  
We have used an independent, large, homogeneously analyzed
sample of NLS1 galaxies (55 objects)
{\em plus} a comparison sample
of BLS1 galaxies (39 objects),
in order to study their locus on
the $M_{\rm BH} - \sigma$ plane. 
We conclude that the width of the [SII] emission line, 
and of the narrow core of the [OIII] line
(but only after removal of [OIII] lines with high blueshifts),
are the best available surrogates for stellar velocity
dispersion $\sigma_*$. Using these lines puts NLS1 galaxies
(and BLS1 galaxies), as a class, on the $M_{\rm BH} - \sigma$ relation
of non-active galaxies.  
Throughout this paper, we use a cosmology with 
$H_{\rm 0}$=70 km\,s$^{-1}$\,Mpc$^{-1}$, $\Omega_{\rm M}$=0.3
and $\Omega_{\rm \Lambda}$=0.7.

\section{Sample selection and data analysis}
We use the sample of NLS1 and BLS1
galaxies of Xu et al. (2007). Sample
selection, data preparation, and data analysis
methods are described in detail in that work.
In brief:  the sample consists of
NLS1 galaxies from the catalogue of Veron-Cetty \& Veron 
(2003), plus a comparison sample of BLS1 galaxies
from Boroson (2003) at $z < 0.3$, which have 
detectable (S/N$>$5) [SII] emission in their
{\it Sloan Digital Sky Survey} (SDSS)
DR3 (Abazajian et al. 2005) spectra. 
The galaxies of the two samples have similar 
redshift and absolute
magnitude distributions. 
Xu et al. (2007) corrected the  SDSS spectra 
for Galactic extinction, decomposed
the continuum 
into stellar and AGN components, and then subtracted
the stellar continuum contribution
and the FeII complexes from the spectra.   
Emission lines were fit with Gaussian profiles.

Special attention was paid to [OIII] profile shapes.
The deviation of [OIII] profiles from single
Gaussians is well known, 
and NLS1 galaxies show strong blue wings
particularly often (Boroson 2005 and ref. therein).
Authors who previously used $\sigma_{\rm [OIII]}$
from NLS1 galaxies were aware of this complication,
and most of them did correct for it.
In most cases they did so by using only the `red' half of
the [OIII] line profile to measure the FWHM.
Wang \& Lu (2001) decomposed the [OIII] profiles
into core and wing.
We followed two approaches in
measuring FWHM$_{\rm [OIII]}$:
(a) determination of the total FWHM of the summed two-component Gaussian
from a careful decomposition into a core and a blue
wing component whenever present, FWHM$_{\rm [OIII]_{\rm totl}}$; 
and (b) the FWHM
of the core component of [OIII], FWHM$_{\rm [OIII]_{\rm core}}$.
Emission lines other than [OIII] are
well represented by a single Gaussian. 
The lines [SII]6716,6731 (hereafter
referred to as [SII]) were both fit with the same width. 
We follow the standard procedure of using 
the radius($R_{\rm BLR}$)-luminosity($L$) relation of BLS1 galaxies
(which appears to hold for NLS1 galaxies; Peterson et al. 2000)  
to compute BH masses, following Kaspi et al. (2005). 
We then compare our results with 
the $M_{\rm BH} - \sigma_*$ relation of Tremaine
et al. (2002; their eqn. 1 and 19) 
and of Ferrarese \& Ford 
(2005, FF05 hereafter; their eqn. 20).
An error in black hole mass of 0.5 dex (as in Grupe \& Mathur 2004)
was conservatively assumed. 
Formal measurement
errors in FWHM are typically 5-20\%.

We note that all SDSS
observations of our sample include emission from the entire 
NLR (so we are not systematically
excluding the outer
parts of the NLR in the 
nearest objects). At the lowest redshift in
our sample, $z$=0.034,
an SDSS fiber diameter of 3$^{\prime\prime}$
corresponds to 2 kpc, larger 
than typical NLRs.

\section{Results}

We investigated how the following factors affect the galaxies' location
on the $M_{\rm BH} - \sigma$ plane:
(a) different methods of fitting the [OIII] profile;
(b) the use of other emission lines
as surrogates for $\sigma_*$; 
and (c) the influence of other parameters, 
like [OIII] blueshift and $L/L_{\rm Edd}$.

\subsection{NLS1 galaxies on the $M_{\rm BH} - \sigma_{\rm [OIII]}$ plane}

First, we confirm that NLS1 galaxies as a class 
deviate from the $M_{\rm BH} - \sigma$ 
relation of non-active galaxies if we use the
full profile of [OIII] to measure $\sigma$ (Fig. 1). 
We computed the deviation 
$\Delta\sigma := \log \sigma_{obs} - \log \sigma_{pred}$
of the NLS1 and BLS1 galaxies from the 
FF05 relation.  
Applying a Kolmogorov-Smirnov (K-S) test shows that 
the two samples are drawn from different parent distributions 
(K-S probability of being drawn from the same population: 0.002). 
NLS1 galaxies follow the $M_{\rm BH} - \sigma$
relation much better
if only the narrow core of [OIII] is used to
measure $\sigma$ (Fig. 1). However, there are still a number
of NLS1 galaxies with very broad [OIII] profiles
and these lie systematically below the $M_{\rm BH} - \sigma$
relation. In Sect. 3.3 we identify the cause of this. 
We show that the majority of these `[OIII] outliers'
are characterized by high NLR outflow velocities,
and therefore, their velocity fields  are not dominated
by the bulge potential and their line widths cannot be used as surrogates
for $\sigma_*$.

\subsection{Widths of other NLR lines as surrogate for $\sigma_*$}

The strongest (high-ionization) emission line in AGN spectra is 
[OIII]5007, which is commonly 
used for $\sigma$ estimates. 
Because of its strength,
we are able to decompose the line into a core and blue wing.  
Low-ionization emission lines generally do not
show this spectral complexity, 
and we therefore also consider using their
widths as surrogates for  $\sigma_*$. 
The strongest lines in the commonly available
wavelength range between $\sim4000-7000$\AA~
are [NII]6584,6548, [SII] and [OI]6300.  [NII]
is, however, blended with H$\alpha$ except in
high-resolution spectra (see Zhou et al. 2006
for the use of $\sigma_{\rm [NII]}$ as a surrogate for $\sigma_*$).
We focus on [SII] here, which is detected in all galaxies of our sample. 
As shown by Greene \& Ho (2005), $\sigma_{\rm [SII]}$ is indeed a good
substitute for $\sigma_*$ in Seyfert 2 galaxies. 

Using the width of [SII] as a surrogate for 
$\sigma_*$, NLS1 galaxies as a class are 
now {\em on} the $M_{\rm BH} - \sigma$ relation (Fig. 1).
Repeating the K-S test 
on the $\Delta\sigma$ distribution of NLS1 and BLS1 galaxies 
(Fig. 3) shows that the
two samples are consistent with being drawn 
from the same parent distribution (K-S probability of 0.4). 
We note that the FWHMs of [SII] and [OIII]$_{\rm core}$ 
correlate well with each other (after removal of `blue outliers';
Sect. 3.3), demonstrating that both emission lines trace similar 
NLR velocity fields (Fig. 1).
Only 40\% of the NLS1 galaxies of our sample have [OI]
detected.
The scatter in $\sigma_{\rm [OI]}$ is relatively small
and most galaxies are close to the FF05 relation,
except two outliers with 
large $\sigma_{\rm [OI]}$.

\subsection{Influence of several parameters  
on the galaxies' location on the $M_{\rm BH} - \sigma$ plane}

We investigated the influence of several 
quantities -- [OIII]$_{\rm core}$ blueshift, $L/L_{\rm Edd}$,
intensity ratio FeII4570/H$\beta$, NLR density, and absolute magnitude 
--  on the
galaxies' location on the $M_{\rm BH} - \sigma$ plane.
We  
divided each of these parameters into 3 bins:
low values, intermediate values, and high values,
such that about equal numbers of NLS1 galaxies
fell in each bin.
We measured the velocity shift of  
[OIII]$_{\rm core}$, $v_{\rm[OIII]}$, 
relative to [SII] (we use positive velocity values for blueshifts). 
$L_{\rm Edd}$ was derived from the black hole mass,
according to $L_{\rm Edd} = 1.3\,10^{38}$ M$_{\rm BH}$/M$_{\odot}$ erg/s. 
The bolometric luminosity $L_{\rm bol}$ was estimated from the luminosity
at 5100\AA, $L_{\rm 5100}$, and a 
bolometric correction of 
$L_{\rm bol}=9\,\lambda{L_{\rm5100}}$. 
The NLR density was derived from the density-sensitive [SII]
intensity ratio,
[SII]6716/[SII]6731  
(Xu et al. 2007), and
the absolute magnitude from the SDSS i-band magnitude. 
Our aim is to search for
trends across the $M_{\rm BH} - \sigma$ plane 
in order to see whether we can identify
physical mechanisms
that determine an object's location
on the plane, and, in particular,
that are responsible for the [OIII] outliers.

First, we find that almost all outliers in $\sigma_{\rm{[OIII]}}$ 
actually show a very  
high blueshift of [OIII]$_{\rm core}$ 
with a velocity shift $v_{\rm[OIII]}$$>$150 km\,s$^{-1}$
[hereafter referred to
as `blue outliers' (Zamanov et al. 2002); marked in Fig. 2].  
More generally, we find the trend across 
the $M_{\rm BH} - \sigma_{\rm{[OIII]}}$ plane that
[OIII] lines of higher $\sigma$ also show 
higher outflow velocity, 
revealing the presence of an extra radial velocity field 
presumably due to outflows.
Among the galaxies with positive $\Delta\sigma$ 
(those below  
the FF05 relation)
49\%  are in
the highest-velocity bin, while only 26\%
are in the lowest bin. 
However, if we exclude all galaxies with
$\sigma_{\rm [OIII]}$ in the highest bin 
($v_{\rm[OIII]}$$>$75 km\,s$^{-1}$), NLS1 and BLS1 galaxies
show similar scatter and agree with the $M_{\rm BH} - \sigma$
relation of normal galaxies (Fig. 3; the K-S test 
shows that the two samples are not
statistically distinguishable with K-S probability 0.5). 
We have checked that [SII] is {\em not} systematically influenced
by the outflow components that appear in [OIII]$_{\rm core}$.
NLS1 galaxies with blue outliers in [OIII]$_{\rm core}$ are uniformly 
distributed across the $M_{\rm BH} - \sigma$ plane in [SII].
 
In order to see whether other parameters affect
an object's location on the $M_{\rm BH} - \sigma_{\rm[OIII]}$ plane
(Fig. 2),
we first removed all objects with the highest
[OIII] outflow velocities ($v_{\rm [OIII]} > 75$ km/s),
and then studied the correlation of each parameter
with the object's deviation from
the FF05 relation, measured in terms of $\Delta\sigma$.  
We cannot identify a strong dependence of $\Delta\sigma$
on $L/L_{\rm Edd}$ (Spearman rank correlation coefficient
$r_{\rm S}$=0.10){\footnote{Note that this does not imply an
inconsistency with Mathur \& Grupe (2005a,b), since their
({\em soft} X-ray selected) sample includes a significant number
of NLS1 galaxies with very high $L/L_{\rm Edd}$, above the highest
value in our sample.}}, 
FeII4570/H$\beta$ ($r_{\rm S}$=0.05),
and density ($r_{\rm S}$=$-$0.01). However, if we keep    
all objects with high blueshifts, $L/L_{\rm Edd}$ and $\Delta\sigma$
are correlated ($r_{\rm S}$=0.34). 
Galaxies
with higher absolute magnitudes M$_{\rm i}$ 
are, on average, located at higher BH masses,
as expected, but there is no systematic trend
horizontally across the plane.

\section{Discussion and conclusions}

We confirm the well-known result (e.g., Grupe 2004)
that NLS1 galaxies
have, on average, lower BH masses than BLS1 galaxies
(in our sample, masses range between
$\log M_{\rm BH, NLS1} = 5.7 - 7.3$ M$_{\odot}$,
while $\log M_{\rm BH, BLS1} = 6.5 - 8.4$ M$_{\odot}$).
We further confirm that NLS1s are off 
the $M_{\rm BH} - \sigma$ relation 
when [OIII]$_{\rm totl}$ is used as surrogate for $\sigma_*$,
and that -- after
removing blue wings -- [OIII]$_{\rm core}$
gives more consistent results.
However, after correction for blue wings several
blue outliers still stick out with very large $\sigma$
values. If these were to reach the $M_{\rm BH} - \sigma_*$
relation moving vertically on
the $M_{\rm BH} - \sigma$ plane by growing 
their black holes, they would have
to increase their current BH masses by
factors $>$100, and become more
massive than the most massive BHs in the BLS1
comparison sample. 
Instead, the fact that all show high outflow
velocities in [OIII]$_{\rm core}$ 
demonstrates that the properties of the NLR
(unusual outflows in the high-ionization NLR)
drive the deviation of these NLS1 galaxies
from the $M_{\rm BH} - \sigma$ relation. 
Those galaxies which are extreme outliers in [OIII]
still follow the $M_{\rm BH} - \sigma$ relation in [SII].

There is good  evidence
that many NLS1 galaxies are accreting at a high rate
relative to the Eddington rate
(even if not as high as in high-redshift quasars;
Warner et al. 2004), so, as emphasized
by 
Mathur \& Grupe (2005a,b) their BHs must be rapidly
growing. 
Since we find that they are on the $M_{\rm BH} - \sigma$  
relation now, their tracks on the
$M_{\rm BH} - \sigma$  plane must
be diagonal, else they would deviate in the future.
We distinguish two possibilities:
(1) NLS1 galaxies as a class {\em evolve}
into BLS1 galaxies  with respect to
their BH mass distribution.
Then, on average, the BHs of NLS1 galaxies 
have to grow by a factor $\sim$10. 
If accreting at $L/L_{\rm Edd}$=1.0  
this  would take them 10$^8$ years.
If they did so, a corresponding change
in $\sigma_*$ of the bulge of the host galaxy
is required in order to ensure
that the $M_{\rm BH} - \sigma_*$  relation still holds. 
Recent host galaxy studies of NLS1s indicate that
they are not preferentially merging systems
(Krongold et al. 2001, Ryan et al. 2007),
but have a relatively high bar fraction
(Crenshaw et al. 2003, Ohta et al. 2007).
In that case, some secular mechanism
must be at work to adjust the bulge properties.
At this stage, we can only
speculate that some of the feedback
mechanisms (Silk \& Rees 1998)
studied
in the context of merger-induced BH fuelling
(e.g., di Matteo et al. 2005, King 2005) 
also hold for bar-induced
fuelling.
Alternatively, (2) the NLS1 mass distribution does {\em not} evolve
into the BLS1 mass distribution. In this picture,
NLS1s are simply low-mass extensions of the 
BLS1 phenomenon (e.g., McHardy et al. 2006).
In that case, the high $L/L_{\rm Edd}$ values could
just represent a relatively {\em short-lived} accretion phase
which would only insignificantly grow
their BHs.
Then, within the large scatter of 
$\sigma$, no adjustment of the host's bulge
would be needed.  
Sources with relatively low $L/L_{\rm Edd}$ could
be before or after an accretion event. 

In summary, we conclude that the NLS1 and BLS1 galaxies
of our sample, as a class, 
follow the $M_{\rm BH} - \sigma_*$ relation
if we use as surrogate for $\sigma_*$ 
the widths of emission lines which are not strongly affected
by outflow components ([SII]; or [OIII]$_{\rm core}$ {\em after} 
removing galaxies  with high outflow velocities $v_{\rm [OIII]}$). 
Further studies of the evolution of highly accreting AGN 
along the $M_{\rm BH} - \sigma_*$ relation  will shed 
new light on the evolution of galaxies in general
and NLS1s in particular. 
\acknowledgments
 We thank  
 D. Merritt, H. Zhou, I. Strateva, 
 S. Mathur, D. Grupe, A. Robinson,
 and G. Hasinger
 for discussions. 
 DX acknowledges 
 the support of the Chinese
 National Science Foundation (NSFC)
 under grant NSFC-10503005.
 This research made use
 of the SDSS data base.

% \clearpage

\begin{figure*}
\plotone{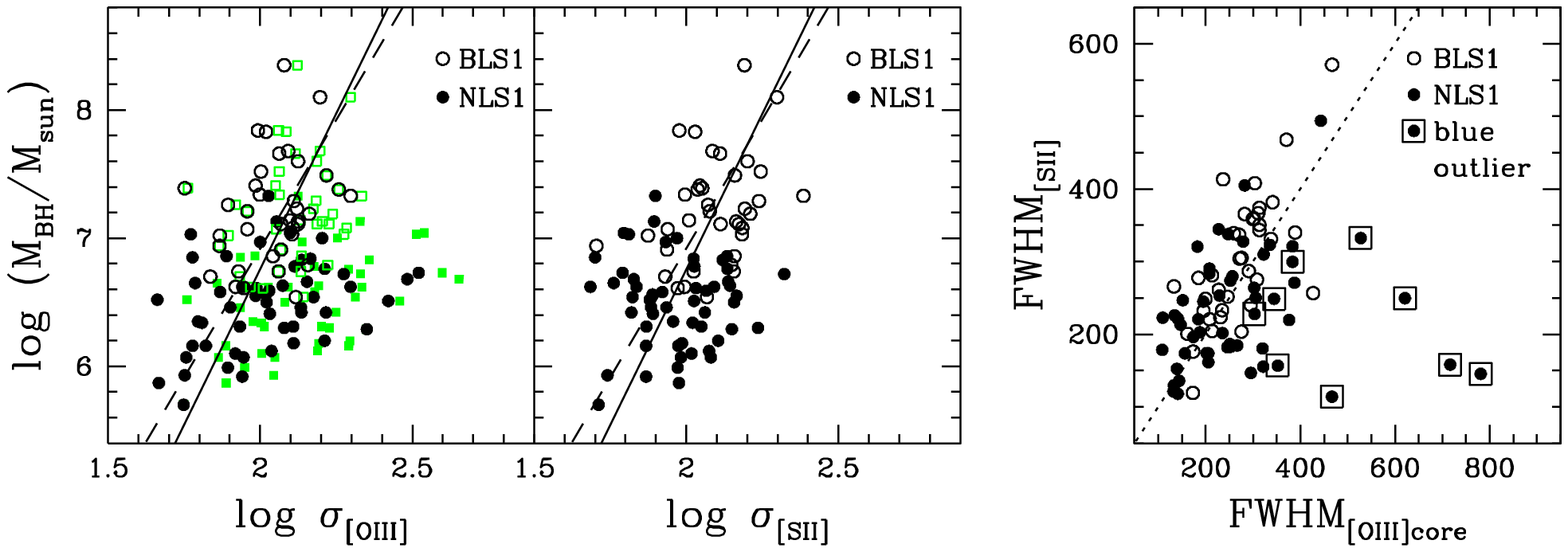}
\caption{ {\bf{Left panels:}} Location of NLS1 
galaxies (filled symbols) and BLS1 galaxies 
(open symbols)
on the $M_{\rm BH}-\sigma_{\rm[OIII]}$ and $M_{\rm BH}-\sigma_{\rm[SII]}$ plane,
respectively.   
In [OIII], 
black circles refer to measurements of the core of the line after correction
for asymmetric blue wings, while green squares correspond to measurements
of the full profile (as described in the text). 
The dashed and solid lines  
correspond to the $M_{\rm BH}-\sigma_*$ relation of non-active
galaxies of Tremaine et al. (2002) and 
of Ferrarese \& Ford (2005),
respectively. $\sigma$ and FWHM are measured in km/s. 
{\bf Right}: Correlation between the FWHM of [SII] and [OIII]$_{\rm core}$
(dotted: one-to-one line fo equal FWHMs -- note the different
scales in x and y axis). 
Blue outliers in [OIII]$_{\rm core}$ ($v_{\rm [OIII]} > 150$ km/s) are marked
with an extra open square.
% This plot demonstrates that the widths of [SII] and [OIII] represent similar 
% NLR velocity fields, with the exception of [OIII] blue outliers. 
}
\end{figure*}

%/clearpage

\begin{figure*}
\plotone{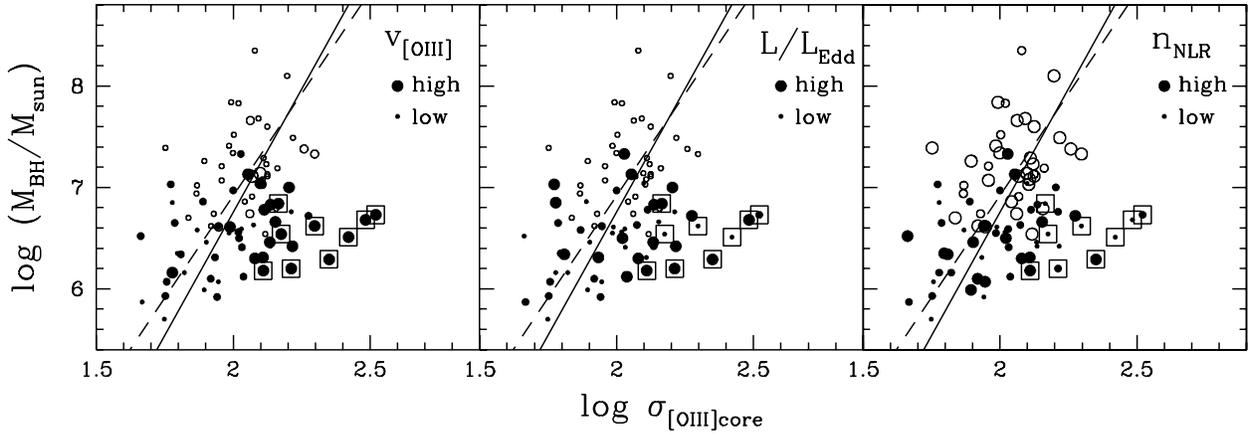}
\caption{Influence of different parameters on 
the galaxies' positions on the $M_{\rm BH}-\sigma_{\rm[OIII]_{\rm core}}$ plane
(open circles: BLS1 galaxies, filled circles: NLS1 galaxies).
Each parameter is devided into 3 bins and coded by circle size:
small: low value, medium: intermediate value,
large: high value. 
From left to right:
coding according to [OIII]$_{\rm core}$ blueshift
($v_{\rm [OIII]}$ $<$30 km/s, 30-75 km/s, $>$75 kms/s in lowest, intermediate, and 
highest velocity bin, respectively),
$L/L_{\rm Edd}$ ($<$0.6, 0.6-0.9, $>$0.9), 
and NLR density ($<$140 cm$^{-3}$, 140-240 cm$^{-3}$, $>$240 cm$^{-3}$).
Blue outliers in [OIII]$_{\rm core}$ are marked 
with an extra open square.  }   
\end{figure*}

\begin{figure*}
\plotone{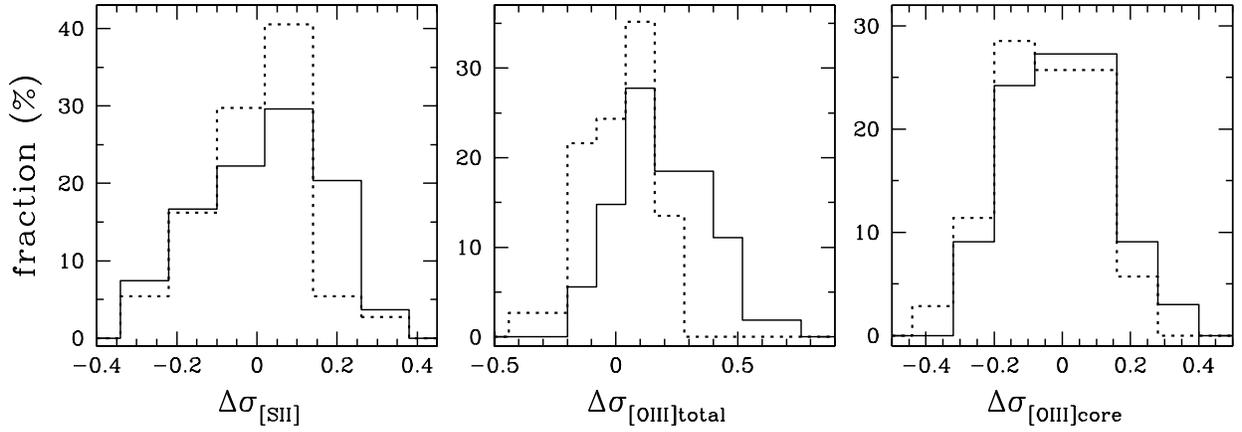}
\caption{Histograms of the deviation $\Delta\sigma := \log \sigma_{obs} - \log \sigma_{pred}$
of galaxies from the $M_{\rm BH} - \sigma$ relation of FF05
(solid line: NLS1 galaxies, dotted line: BLS1 galaxies) 
for [OIII]$_{\rm core}$
({\em after} excluding objects in the highest velocity bin), [OIII]$_{\rm totl}$,
and [SII], 
from right to left.    
}
\end{figure*}

%\clearpage

\end{document}